# The Information Service Evaluation (ISE) Model

**Laura Schumann**
Dept. of Information Science, Heinrich Heine University Düsseldorf, Germany.
E-mail: laura.schumann (at) hhu.de

**Wolfgang G. Stock**
Dept. of Information Science, Heinrich Heine University Düsseldorf, Germany.
E-mail: stock (at) phil.hhu.de

## Abstract

Information services are an inherent part of our everyday life. Especially since ubiquitous cities are being developed all over the world their number is increasing even faster. They aim at facilitating the production of information and the access to the needed information and are supposed to make life easier. Until today many different evaluation models (among others, TAM, TAM 2, TAM 3, UTAUT and MATH) have been developed to measure the quality and acceptance of these services. Still, they only consider subareas of the whole concept that represents an information service. As a holistic and comprehensive approach, the ISE Model studies five dimensions that influence adoption, use, impact and diffusion of the information service: information service quality, information user, information acceptance, information environment and time. All these aspects have a great impact on the final grading and of the success (or failure) of the service. Our model combines approaches, which study subjective impressions of users (e.g., the perceived service quality), and user-independent, more objective approaches (e.g., the degree of gamification of a system). Furthermore, we adopt results of network economics, especially the "Success breeds success"-principle.

## Keywords

Information services; Technology acceptance; Evaluation

**Introduction**

### 1. Information needs, information services and their appropriate evaluation

Complex information services satisfy complex human information needs. Information needs find their expression in human information behavior including the behavior of information production (e.g., user-generated content in social media) and the behavior of information seeking (e.g., browsing through web sites or applying search engines). Complex information services are, for instance, governmental websites of states or cities (Almalki, Duan, & Frommholz, 2013; Mainka et al., 2013), Web 2.0 services like Facebook, YouTube, Flickr and Twitter (Lin & Lu, 2011), mobile services (López-Nicolás, Molina-Castillo, & Bouwman, 2008) or city-wide digital services in ubiquitous cities (which is one of our examples; Schumann, Rölike, & Stock, 2013). The construction and maintenance of complex information services is expensive. Heeks (2003, p. 2) reports that only 15% of all e-government projects in developing or transitional countries succeed in contrast to 35% which are total failures and 50% which are partial failures. That is why we are in need of identifying successful services. How do users adopt, use and accept those services? Do the information services exercise influence over the users' behavior? How do such services diffuse into society? Our research question is: How can we evaluate even large and complex information systems in the sense of their adoption, use, impact and diffusion? Kusunoki and Sarcevic (2013, p. 860), on the iConference 2013 in Fort Worth, TX, describe this problem accurately:

> Users and the information systems designed to support their needs and behaviors are becoming increasingly complex. Evaluators are tasked with designing evaluation methods that address the evaluation challenge of systems conceived through newer design principles, while also identifying issues and user perceptions in an efficient and effective manner.

Starting point is the user. He or she will adopt, use and accept an information service – or will reject it. But he or she will only accept a service, if it pays off the user and satisfies his or her information need. So what is a human information need? According to Maslow (1954), the fundamental needs of humans are breathing, food, water, sex and sleep. Further needs such as safety, love, esteem and self-actualization are based on these physiological needs. Information is not a part of them. The need for information arises when one of the human needs cannot be satisfied right away. If such a situation is given we start to produce information or to look for information that will help us to satisfy our need. Wilson (1981, p. 8) describes the latter situation as "information-seeking towards the satisfaction of needs" which can – in combination with the need of producing information – be considered as "information need".

For this reason information behavior has always been a part of the daily human life (Spink, 2010). Nowadays we tend to base almost all our decisions on information gathered in the World Wide Web or elsewhere. Furthermore Case (2007, p. 18) states:

> "Every day of our lives we engage in some activity that might be called information seeking, though we may not think of it that way at the time. From the moment of our birth we are prompted by our environment and our motivations to seek out information that will help us meet our needs."

Especially applying social media, users generate their own content and publish it. Possibly even they sometimes index their pieces of information with content-describing words, called "tags" (Peters, 2009).

So information simplifies or improves the human life in many different ways and influences it all the time. In the majority of cases we are not able to satisfy the arising need in that moment, hence an information need emerges. To facilitate the information production and seeking processes, several information services have been developed over time to enable the people to publish and to access the needed information.

Lately the number of these information services has been constantly growing. This increase is due to the role of computers in our everyday life. With the consistent further development of computers and information and communication technology (ICT) the way of exchanging information and thereby the information services have changed. Röcker (2010) found that the existing evaluation models have to be adapted to be able to measure the new generation of information services. There are new aspects that have to be taken into account such as the acceptance amongst the users, e-governance and the culture the information service is integrated into. The existing models are limited in their scope since they only focus on certain aspects of the big picture of information services.

Nowadays the conditions the information services are based upon are different because almost every user has at least one personal computer as well as one smartphone; and ICT services are invisibly embedded in everyday objects to make our lives easier and more comfortable by offering the possibility to publish information or to access the needed information in the moment it is required no matter where we are. This integration of ubiquitous computing into our everyday life has first been mentioned by Mark Weiser (1991).

## 2. An example: Information services in u-cities

For Weiser (1993), ubiquitous computing is "the idea of integrating computers seamlessly into the world" (1991, p. 94). In a ubiquitous city (or, in short, a u-city), ubiquitous computing is realized on city-level. Information is omnipresent and everyone should be able to create and to retrieve information whenever and wherever a need arises.

You can find u-cities and approaches to construct them all over the world. Especially in Korea there are lots of projects of such information-rich cities (Shin, 2009; Shin, 2010; Lee et al., 2008), but there is a u-city in Europe as well, namely Oulu in Finland (Schumann, Rölike, & Stock, 2013). U-city services consist, for instance, of services delivered via media poles such as the touch screen monoliths in Seoul's Gangnam district or in the city center in Oulu, of services created for the use of smartphones (apps), and of services which depend on sensors (Figure 1). The services are oriented on the city-region and are context-aware (with regard to the user and the place and time she or he stays). The services are pull services (when the user is asking the system) as well as push services (when the system actively informs the user). There are specific services for and by companies, administrations, citizens and other user groups, e.g., tourists. "A u-city ... includes a sensory network and context-ware information management systems with a variety of distributed devices and autonomously working software" (Kwon & Kim, 2007, p. 151). If the city additionally refers to sustainability and livability, some authors speak of "smart cities" (Hollands, 2008; Chourabi et al., 2012). In context with other infrastructures (for the knowledge city, the creative city and the green city) u-city services form groundwork for emerging cities in the knowledge society, the so-called "informational cities" (Castells, 1989; Stock, 2011; Mainka, Khveshchanka, & Stock, 2011).

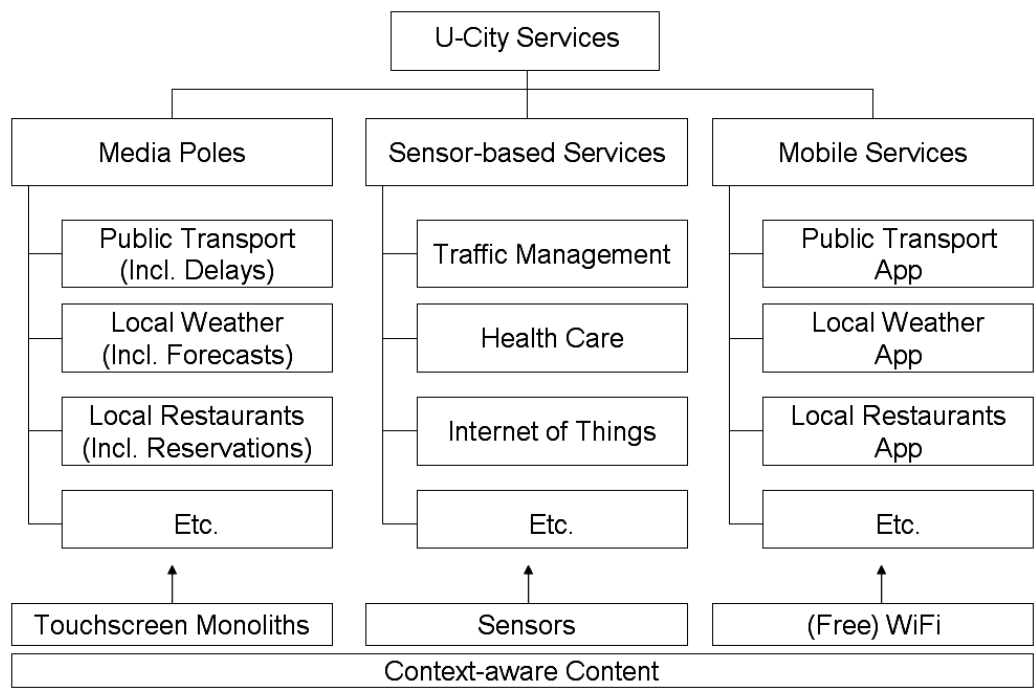

Figure 1. Exemplary Services of a Ubiquitous City.

In the light of the complexity of the u-city services, it becomes clear that it is not possible to apply the classical Technology Acceptance Model (TAM) (Davis, 1989). So, for instance, it does not make much sense to ask users for the ease of use of free WiFi. And TAM has been applied to study acceptance of information systems in companies. When we analyze u-city services the scope must be much broader: We have to consider all members of a society, including children, students, households, social communities, etc. "This is also attributable to Information Technology becoming a ubiquitous part of daily life ever since the introduction of ICTs" (Choudrie, Olla, & Bygstad, 2010, p. i).

## Models and Techniques to Evaluate Information Services and Their Acceptance

We introduce a holistic and comprehensive model that allows us to span a theoretical framework for all aspects of the evaluation of (even large) information services. It is valid for the study of information services in companies and other institutions as well as in everyday life, for instance in households, in schools or in social communities. We try to integrate advantages of evaluation and information acceptance models from different scientific fields, including information systems research, marketing research, knowledge management, software engineering, computer science, and information science. Among others, we studied the following models, measures, instruments and constructs (sorted by the time of publication of the seed article):

- Critical Incident Technique (Flanagan, 1954),
- Effectiveness Models of Information Retrieval (Recall and Precision) (Kent, Berry, Luehrs, & Perry, 1955),
- SERVQUAL (Parasuraman, Zeithaml, & Berry, 1988),
- Technology Acceptance Model (TAM) (Davis, 1989),
- DeLone & McLean Model (DeLone & McLean, 1992; revised: 2003),
- Usability (Nielsen, 1993),
- IT SERVQUAL (Pit, Watson, & Kavan, 1995),
- Sequential Incident Technique (Stauss & Weinlich, 1997),
- TAM 2 (Venkatesh & Davis, 2000),
- Unified Theory of Acceptance and Use of Technology (UTAUT) (Venkatesh, Morris, Davis, & Davis, 2003),
- Model of Adoption of Technology in Households (MATH) (Brown & Venkatesh, 2005),
- Jennex & Olfman Model (Jennex & Olfman, 2006),
- Customer Value Research (McKnight, 2006),
- TAM 3 (Venkatesh & Bala, 2008).

All analyzed articles study important aspects of information services. Some models were constructed to study the acceptance of information services in companies and other organizations (TAM, TAM 2, TAM 3, UTAUT, DeLone & McLean Model, Jennex & Olfman Model), one model was developed to evaluate the acceptance of information services in households and other everyday situations (MATH), and the rest of the models can be applied in both, the business and everyday context as well.

A historical point of origin for the evaluation of the quality of information systems in the business area is the registration of technology acceptance in the workplace. The Technology Acceptance Model (TAM) (Davis, 1989) uses subdimensions (initially: perceived ease of use and perceived usefulness) in order to measure the quality of an information service's technical make-up. In TAM 2, Venkatesh and Davis (2000) showed that perceived usefulness is dependent on other factors including the user's experience, voluntariness, social influences (called "subjective norm"), image, output quality in relation to the job and result demonstrability. Perceived ease of use correlates with control (computer self-efficacy and facilitating conditions), with the intrinsic motivation of the user and with his/her emotions (Venkatesh, 2000). The construction of Technology Acceptance Models climaxed with the Unified Theory of Acceptance and Use of Technology (UTAUT) (Venkatesh, Morris, Davis, & Davis, 2003). Here, four user-specific criteria (gender, age, experience and voluntariness of use) meet four aspects of the user-system relationship (performance expectancy, effort expectancy, social influence and facilitating conditions). Performance expectancy includes the well-known perceived usefulness and effort expectancy the perceived ease of use. The two other aspects are known from TAM2.

TAM, TAM2 and UTAUT find their applications in business contexts. On the example of the adoption of personal computers in homes (Venkatesh & Brown, 2001), Brown and Venkatesh (2005) constructed their Model of Adoption of Technology in Households (MATH). MATH works with a set a users' beliefs and includes attitudinal beliefs (e.g., application for personal use, utility for children or status gains), normative beliefs (among others, friends and family influences as well as influences from TV, newspaper, etc.) and control beliefs (costs, ease of use, requisite knowledge).

Venkatesh (2000) conceptualized intrinsic motivation as computer playfulness. With the development of the World Wide Web (Moon & Kim, 2001), of digital games – or "pleasure-oriented (or hedonic) information systems" (van der Heijden, 2004) and of services of the Web 2.0 (Knautz, Soubusta, & Stock, 2010) the dimension of perceived fun as a result of perceived playfulness (Lieberman, 1977; Barnett, 1990) became an important building block of the perceived information system quality. Especially with the successful implementation of e-commerce systems, a further dimension emerged: perceived trust (Gefen, Karahanna, & Straub, 2003).

Meta-analyses of TAM (Legris, Ingham, & Collerette, 2003; King & He, 2006) show the usefulness of this model (in organizational settings as well as in household, residential and consumer contexts; Dwivedi et al., 2010), but they show also, that TAM has to be integrated into a broader model. For different user groups (students, professional users and general users; King & He, 2006, p. 748) and for different tasks (job-office applications, general applications and e-commerce and internet applications; King & He, 2006, p. 749) the effects measured by TAM differ widely. Especially in environments, where information services are ubiquitously available (Röcker, 2010), TAM, UTAUT and MATH only reflect parts of the whole story, insofar they limit themselves on the technology of the service under study.

In the model proposed by DeLone and McLean (1992), the technical dimension is joined by that of information quality. Insofar information services depend on content (and most do so), we have to regard this aspect. The perceived content quality concentrates on the knowledge that is stored in the system.

DeLone and McLean (2003) as well as Jennex and Olfman (2006) expand the model via the dimension of service quality. When analyzing perceived service quality, the objective is to inspect the services offered by the information system and the way they are perceived by the users. To study service quality, there are "classical" techniques such as the critical incident technique (Flanagan, 1954) or the sequential incident technique (Stauss & Weinlich, 1997) for the analysis of the whole service process; and SERVQUAL (Parasuraman, Zeithaml, & Berry, 1988) or IT SERVQUAL (Pit, Watson, & Kavan, 1995) for the analysis of attributes of the service. While SERVQUAL measures expectations and experiences of the services' users, Customer Value Research (McKnight, 2006) works with the experience values of the users and with the expectation values of the service developers, which leads to an expression of "irritation", i.e., the misunderstandings between the developers of an IT service and their customers.

The quality of an information service depends not only on the perception of its quality by the users, but also upon objective (user-independent) measures of the service's quality. Aspects of objective service quality include the range of functions it offers (Stock & Stock, 2013, pp. 486-488), its usability (Nielsen, 1993), and the system's effectiveness (offering the "right" services; Drucker, 1963) and efficiency (touch screen sensibility and speed of system reactions, amongst others). For instance, in information retrieval systems (search engines on the WWW and commercial research systems in the Deep Web), efficiency measures how quickly a search will be processed, and effectiveness the ability of the system to find the right information (and only the right information) (Croft, Metzler, & Strohman, 2010, p. 297). The classical indicators of retrieval system's effectiveness are recall and precision (Kent, Berry, Luehr, & Perry, 1955). To perform objective studies one can work with analyses of log files, with user surveys, and with systematic observations of test users in a laboratory setting or in a real-life situation.

## The Information Service Evaluation (ISE) Model

Our Information Service Evaluation (ISE) Model (Figure 2) consists of five dimensions:
- Information service quality,
- Information user,
- Information acceptance,
- Information environment, and
- Time.

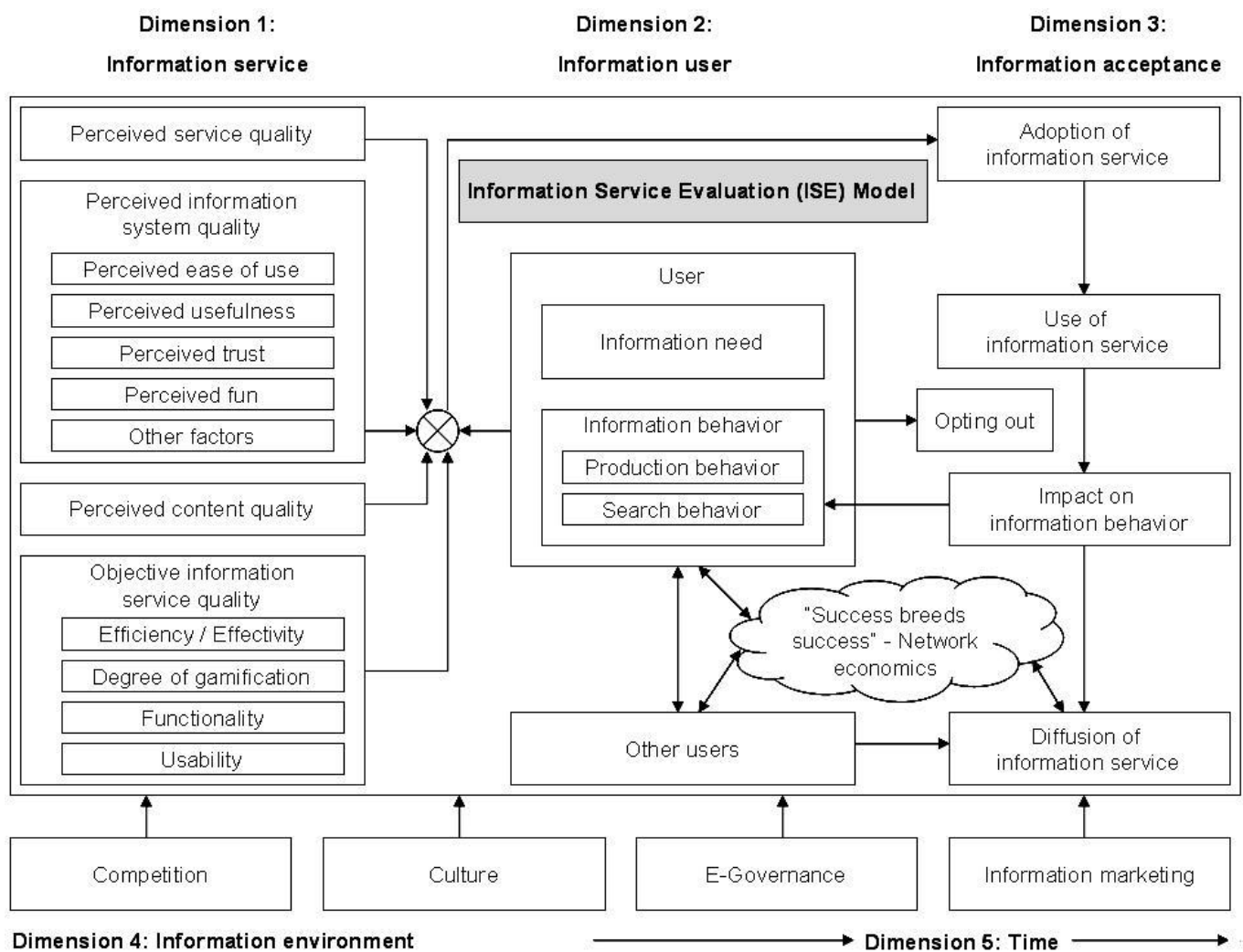

Figure 2. The Information Service Evaluation (ISE) Model

# 1. Information service quality

The quality of an information service can be analytical divided into the perceived service quality (the information service quality as a user estimates it) and the "objective" information service quality (as an expert with scientific concepts will describe it). The user-oriented quality estimation can be divided into three dimensions:

- Perceived service quality,
- Perceived information system quality (ease of use, usefulness, trust, fun and other factors), and
- Perceived content quality.

Additionally, we work with aspects to get an objective impression of the service's quality:

- Efficiency,
- Effectiveness,
- Functionality,
- Degree of gamification and
- Usability.

## 1.1. Perceived service quality

For registering the process component of an IT service we apply the sequential incident technique and the critical incident technique. In the sequential incident technique (Stauss & Weinlich, 1997), users are observed while working through the service in question. Every step of the process is documented, which produces a "line of visibility" of all service processes – i.e., displaying the service-creating steps that are visible to the user. If the visible process steps are known, users can be asked to describe them individually. This is the critical incident technique (Flanagan, 1954). Typical questions posed to users are "What would you say is the primary purpose of X?" and "In a few words, how would you summarize the general aim of X?"

For evaluating the attributes of services we use SERVQUAL (Parasuraman, Zeithaml, & Berry, 1988). SERVQUAL works with two sets of statements: those that are used to measure expectations about a service category in general (EX) and those that measure perceptions (PE) about the category of a particular service. Each statement is accompanied by a seven-point scale ranging from "strongly disagree" (1) to "strongly agree" (7). For the expectation value, one might note that "In retrieval systems it is useful to use parentheses when formulating queries" and ask the test subject to express this numerically on the given scale. The corresponding statement when registering the perception value would then be "In the retrieval system X, the use of parentheses is useful when formulating queries." Here, too, the subject specifies a numerical value. For each item, a difference score $Q = PE - EX$ is defined. If, for instance, a test subject specifies a value of 1 for perception after having noted a 4 for expectation, the Q value for system X with regard to the attribute in question will be $1 - 4 = -3$.

Parasuraman, Zeithaml and Berry (1988) define five service quality dimensions (tangibles, reliability, responsiveness, assurance, and empathy). This assessment is conceptualized as a gap between expectation and perception. It is possible to adopt SERVQUAL for measuring the effectiveness of information systems (Pitt, Watson, & Kavan, 1995). In IT SERVQUAL, there are problems concerning the exclusive use of the difference score and the pre-defined five quality dimensions. It is thus possible to define separate quality dimensions that are more accurate in answering specific research questions than the pre-defined dimensions. The separate dimensions can be derived on the basis of the critical processes that were recognized via sequential and critical incident techniques. It was suggested to not only apply the difference score, but to add the score for perceived quality, called SERVPERF (Kettinger & Lee, 1997), or to work exclusively with the perceived performance scoring approach. We work with three scales, the expectation values, the perception values, and the differences between expectation and perception values. If a sufficient amount of users were used as test subjects, and if their votes were, on average, close to uniform, SERVQUAL would seem to be a valuable tool for measuring the quality of IT systems' attributes.

## 1.2. Perceived system quality

The dimension of perceived system quality is the playground of most of the models such as TAM, TAM2, UTAUT and MATH. In our studies, we apply perceived ease of use, perceived usefulness, perceived trust, perceived fun and a residue class "other factors" (including, for instance, perceived costs).

When evaluating perceived IT system quality, questionnaires are used. The test subjects must be familiar with the system in order to make correct assessments. For each subdimension, a set of statements is formulated that the user must estimate on a 7-point scale (from "extremely likely" to "extremely unlikely"). Davis (1989, 340), for instance, posited "using system X in my job would enable me to accomplish tasks more quickly" to measure perceived usefulness, or "my interaction with system X would be clear and understandable" for the aspect of perceived ease of use. In addition to the five subdimensions, it must be asked if and how the test subjects make use of the information system. (This question is important for the aspect of use in dimension 3: information acceptance, too). If one asks factual users (e.g., citizens in a u-city who often use the touch screen monoliths), estimates will be fairly realistic. A typical statement with regard to registering usage is "I generally use the system when the task requires it." It is useful to calculate how the usage values correlate with the values of the subdimensions (and how the latter correlate with one another). A subdimension's importance rises in proportion to its correlation with usage.

## 1.3. Perceived content quality

The quality of the content that is depicted in an information service can vary significantly, depending on the information service analyzed (Stvilia, Gasser, Twidale, & Smith, 2007).

Information services on scientific-technological literature (such as the ACM Digital Library) contain scientific articles whose quality has generally already been checked during the publication process. This is not the case in Web search engines. Web pages or documents in sharing services (e.g., videos on YouTube) are not subject to any process of evaluation. The content quality of such documents is extremely hard to quantify. Here we might ask users for aspects such as freshness of content, its believability, objectivity, readability or understandability (Parker, Moleshe, De la Harpe, & Wills, 2006).

## 1.4. Objective information service quality

"Objective" in this context means that the measurement results are not based solely on users' perceptions, but – wherever it is possible – on other approaches that work independently from end user estimates. The efficiency measure orientates on "doing things right." First of all, for information system this means doing the job as fast as possible. Depending on the system, there are further efficiency criteria, such as the sensibility and error-proneness of touch screens of the media pillars or the availability of high-speed broadband WiFi in all regions of a city. According to Drucker (1963), effectiveness means "doing the right things." Sometimes, effectiveness of an information service is hard to quantify. For all kinds of retrieval systems, however, we apply recall and precision (and – for search results ranked by relevance – Mean Average Precision; Croft, Metzler, & Strohman, 2010, p. 313) to evaluate their effectiveness. Paradigms for our studies are the Cranfield tests (Cleverdon, 1967) and the Text Retrieval Conferences (TReC) (Voorhees, 2002; Harman; 2011). The measure of functionality of an information service is the extent of its functions for information production and information searching (measured independently from the factual application by the users).

Usable information services are those that do not frustrate the users. A common procedure in usability tests in accord with Nielson (1993) is task-based testing. Here an examiner defines representative tasks that can be performed using the system and which are typical for such systems. Such a task for evaluating the usability of a search engine might be "Look for documents that contain your search arguments verbatim!" Test subjects should be "a representative sample of end users" (Rubin & Chisnell, 2008, p. 25). The test subjects are presented with the tasks and are observed by the examiner while they perform them. For instance, one can count the links that a user needs in order to fulfil a task (in the example: the number of links between the search engine's homepage to the verbatim setting). An important aspect is the difference between the shortest possible path to the target and the actual number of clicks needed to get there. The greater this difference is, the less usable the corresponding system function will be. An important role is played by the test users' abandonment of search tasks ("can't find it"). Click data and abandonment frequencies are indicators for the quality of the navigation system (Röttger & Stock, 2003). It is useful to have test subjects speak their thoughts when performing the tasks ("thinking aloud"). The tests are documented via videotaping. Use of eye-tracking methods provides information on which areas of the screen the user concentrated on (thus possibly overlooking a link). In addition to the task-based tests, it is useful for the examiner to interview the subjects on the system (e.g., on their overall impression of the system, on screen design, navigation, or performance). Benchmarks for our usability tests are generally set at a minimum of ten test subjects and a corresponding number of at least ten representative tasks.

Some information systems adopt elements of gamification, i.e., the use of game mechanics in non-game contexts, to motivate the users to continue using the system (Zichermann & Cunningham, 2011). Game mechanics consist of point systems, levels, challenges, virtual goods, leaderboards, gifting and charity and – very important – quests (Knautz, Göretz, & Wintermeyer, 2014). Quests are answered by single persons or collectively by groups. In this way, game mechanics lead to close bonds between players and between the system and the users. Under certain conditions, the user has the experience of "flow" (Czíkszentmihályi, 1975), which means that she or he is engrossed with the system and loses awareness of other things. We describe the system's degree of gamification by counting applied game mechanics.

## 2. Information user

Information science separates three groups of users (Stock & Stock, 2013, pp. 467-468). An information professional is an expert in working with information systems. The professional end user is a specialist in an institution who processes, ad hoc, simple information needs that arise in the workplace. The layman end user, finally, usually is applying search engines, some Web 2.0 services (e.g., Facebook and YouTube), email and perhaps some digital games. Depending on the level of their information literacy (Stock & Stock, 2013, Ch. A.5), users from the three groups will interact in different ways with information services. Additionally, it is necessary to analyze the degree of the user's knowledge of the particular service. If we, for instance, are going to evaluate a specific weblog service, we have to understand the test users' experiences with weblogs in general and with the specific service (Li & Kishore, 2006).

A central point for using or non-using an information service is the information need of a person (Kim, Kim, & Kim, 2010). An individual's information need is the starting point of any information behavior (information production as well as information search behavior).

In addition to the user's information behavior and his or her information needs, we have to study further person-related factors such as

- gender (e.g., Evans, Hopper, Jones, & Knezek, 2013),
- age (e.g., Birkland, 2009),
- digital native / digital immigrant (Prensky, 2001; Agosto & Abbas, 2010),
- culture, nationality, ethnic identification (e.g., Ayouby, Croteau, & Raymond, 2013).

Without user research, no serious evaluation of information services is possible (Herold, 2010).

### 3. Information acceptance

We consider information acceptance as a concept consisting of the aspects adoption, usage, impact and diffusion. If the "right" person in an appropriate situation meets the "right" information service, she or he will adopt and use this service. Adoption has two faces. Supplier-side adoption means that a service-provider realizes a certain service. If the company is the first one which introduces this service, it is a "first-mover" or an "innovator"; if not, it is a "second-mover" or an "imitator" (Linde & Stock, 2011, pp. 361-375). In contrast, consumer-side adoption means that a user applies the supplied service for the first time (Dwivedi et al., 2008).

Adoption does not mean use. One can adopt a service and stop to use it. And one can adopt it and use it permanently. We speak of use, when the user applies some of the information service's functionalities in his or her professional or private life when there is an information need on hand. We know from empirical investigations (Jasperson, Carter, & Zmud, 2005) that most users of information services only apply a narrow band of features and operate at low levels of feature use – but this behavior counts as use as well. Users are frequently concerned to integrate an information service (or this tiny part of it she or he really uses) into their preexisting lifestyle and attendant habits and usages (Herold, 2010). The transition from adoption to use – the continuance – is guided by user satisfaction and perceived usefulness (Bhattacherjee, 2001).

In the case of use it is possible that the user's information behavior will change. This aspect we will call impact. A good example of impact of an information service is Facebook. Just a decade ago, no one spent time on social networks. Today, more than 60% of young students in Germany (about 11 years old) apply social networks at least once a day, spending more than 1 hour per day on average with reading and writing posts (Orszullok, 2013, p. 93). Older students (about 17 years old) use Facebook still more: here the figures are 84%, who uses Facebook daily or more often, and they work within their social network about 2 hours a day (Förster, 2013, p. 130).

Finally, an information service will diffuse into a society, when many people use it and it has impact on their information behavior. Here we find again the aspect of social influences from TAM2 and UTAUT. Diffusion is a typical phenomenon of network economics following the principle of "success breeds success." The more users an information service is able to attract the more the value of the service will increase. More valuable services will attract further users. If an information service passes the critical mass of users, network effects will start. This leads to positive feedback loops for direct network effects (more users – more valuable service – any more users) and indirect network effects (more complementary products – more valuable service – any more complementary products) and – when indicated (Weitzel, Beimborn, & König, 2006) – in the end to a standard (Linde & Stock, 2011, pp. 51-61). Diffusion is a social process depending on the extent to which friends, family members, peers, colleagues, club members, etc. influence a user's information behavior (Niehaves, Gorbacheva, & Plattfaut, 2012). Furthermore arguments pro or against an information service as well as the rhetorical competence of the speakers seem to play roles in the process of information services' diffusion (Barrett, Heracleous, & Walsham, 2013). So it is important to analyze not only the arguments, but also the speakers, their role in society and their rhetorical talents.

Researchers may not forget the aspect of quitting an information service. Opting-out or "pushback" (Morrison & Gomez, 2014) is the result of resistance to (too) frequent usages of services (e.g., of social networks), of perceived breaches of privacy, of over-touching by the service (e.g., too many meaningless posts on Facebook or Twitter), etc. In such a case, users see a negative influence on their (information) behavior and respond accordingly.

We study information acceptance by using questionnaires with typical questions concerning adoption, usage, impact and diffusion, not to forget the relations to other users. For instance, questions with regard to other people and diffusion are, "Would you recommend the information service to other people?" and "Are you influenced in the choice of the information service by other people?" A second way to get information on information acceptance is to interview important stakeholders of the information service.

## 4. Information environment

Information services and information users are embedded in contexts. Important aspects of the information environment are cultural influences (Ayouby, Croteau, & Raymond, 2013), governance (Yates, Gulati, & Weiss, 2013), the market situation (including competitive services) and marketing for the information services (van den Berg & van Winden, 2002). For instance, for mobile broadband diffusion it seems to be essential that countries encourage competition in the market and practice sound regulation (Yates, Gulati, & Weiss, 2013).

To study the environment of information services we use published literature, company reports, political programs, etc. as well as in-depth interviews with stakeholders of the project. When applying conversations, we make use of semi-structured interviews (with an interview guideline) since it offers the possibility to go into detail if necessary. If we study location-critical information services (like u-city systems) we perform – as a matter of principle – ethnographic fieldwork on-site (Brewer, 2000).

## 5. Time

Information services have their own history. A very new service shows different adoption, usage, impact and diffusion figures than a well-established service. So it is necessary to conduct longitudinal studies and to evaluate an information service in the course of time (Venkatesh & Davis, 2000). One cannot compare the adoption, usage, impact and diffusion figures of a new service with the ones of a well-established service that existed already over a number of years.

## Conclusion

The literature discussed in the introduction reveals that new evaluation models for measuring today's information services have to be developed since the characteristics of these services have changed over time. This variance is due to the consequent technological advance we face nowadays. Additionally the importance of information services has run up and their development is often supported by governments that have recognized the importance of offering high quality information services in the different areas of our lives.

The ISE Model with its five different dimensions information service quality, information user, information acceptance, information environment and time is such a model that suits the characteristics of today's information services. It can be used to study the power of information services and, as this paper shows, it is not limited to a specific kind of information service (e.g., search engines). It can be applied to many different kinds of information services such as interactive touch screens with city-specific content or different search engines and it offers a wide choice of aspects that can be evaluated. This way, it is also possible to apply only some of the characteristics shown in the model. It is suitable for the evaluation of user-centered aspects as well for measuring system performance criteria as the appliance in the examples given above shows. Therefore the ISE Model can be used in every information-related discipline (e.g., information systems research, computer science, information science, library science, applied social studies).

The ISE Model has been applied in several different studies to measure the quality and acceptance of information services. We worked with the model on a large project of u-city services (Schumann, Rölike, & Stock, 2013), on the evaluation of a game-based learning platform for higher education (Orszullok & Knautz, 2014) and on evaluation projects of specialized Web-based search engines (Knautz, Soubusta, & Stock, 2010; Knautz, Siebenlist, & Stock, 2010). Especially the example of the u-city of Oulu, Finland, reveals, that even surveys with a high amount of users and very complex information services like they are applied in ubiquitous cities can be examined by the use of the ISE Model. Besides such anecdotic evidence on the success of the ISE model, there is no systematic evaluation of the model, which should be a task for future work. Till then, we believe in the truth of the proverb, "the proof of pudding is in the eating." The proof of an evaluation model is in its successful application in practice.